\documentclass{aa}
\usepackage[varg]{txfonts}
\usepackage{bm}
\usepackage{color}
\usepackage{natbib}
\usepackage{amssymb}
\usepackage{ulem}
\usepackage{etoolbox}
\usepackage{orcidlink}

\bibpunct{(}{)}{;}{a}{}{,} 

\newcommand{\rhog}{{\rho_{\rm g}}}

\newcommand{\vg}{{\bf{v}_{\rm g}}}

\newcommand{\vgx}{{v_{{\rm g}x}}}
\newcommand{\vgy}{{v_{{\rm g}y}}}
\newcommand{\vgz}{{v_{{\rm g}z}}}

\newcommand{\ud}{{\bf{u}}}

\newcommand{\rmd }{{\rm d}}  

\newcommand{\taus}{{\tau_{\rm s}}}

\usepackage{natbib}
\bibpunct{(}{)}{;}{a}{}{,} 

\usepackage{hyperref}

\begin{document}

\title{Nonlinear Evolution of the Unstratified Polydisperse Dust Settling Instability}
\titlerunning{Unstratified Polydisperse Settling Instability} 

\author{Hossam Aly\orcidlink{0000-0002-1342-1694}
          \inst{1}\fnmsep\thanks{E-mail: hossam.saed@gmail.com}
          \and
          Sijme-Jan Paardekooper\orcidlink{0000-0002-8378-7608}\inst{1}
          }
\authorrunning{H. Aly \& S.-J. Paaredekooper}

\institute{Faculty of Aerospace Engineering, Delft University of Technology, Kluyverweg 1, 2629 HS Delft, The Netherlands\\
            \email{hossam.saed@gmail.com}
             }

\date{Received XXX, 2024; accepted YYYY, 20ZZ}

 
\abstract
   {The Dust Settling Instabilty (DSI) is a member of the Resonant Drag Instabilities (RDI) family, and is thus related to the Streaming Instability (SI). Linear calculations found that the unstratified monodisperse DSI has growth rates much higher than the SI even with lower initial dust to gas ratios. However, recent nonlinear investigation found no evidence of strong dust clumping at the saturation level. }
   {To investigate the nonlinear saturation of the mono- and polydisperse DSI. We examine the convergence behaviour wrt. both the numerical resolution as well as the number of species. By characterising the morphology of the dust evolution triggered by the DSI, we can shed more light on its role in planetesimal formation.}
   {We perform a suite of 2D shearing box hydrodynamic simulations with the code \textsc{Idefix}, both in the mono- and polydisperse regimes. We focus on the time evolution of the maximum dust density, noting the time at which the instability is triggered, as well as analyse the morphology of the resultant structure.}
   {In our monodisperse DSI simulations, the maximum dust density increases and the instability saturates earlier with higher spatial resolution, with no signs of convergence yet. The polydisperse simulations do seem to converge with the number of species and produce maximum dust densities that are comparable to, albeit lower than, the monodisperse simulations. Different dust species tend to form adjacent but separate dust filaments, which may have implications on dust growth and further clumping. }
   {The monodisperse DSI produces dust structure at densities high enough that likely leads to clumping. The polydisperse DSI produces lower but comparable dust densities at the same spatial resolution. Our idealised treatment suggests that the DSI is important for planetesimal formation, as it suffers less than the SI from including a dust size distribution. }


\maketitle
\section{Introduction}

Planet formation theory requires dust to grow in proto-planetary discs from micron-sized particles, as inherited from the interstellar medium, to km-sized planetesimals in less than 10 Myr. Dust growth by coagulation is efficient for small particles, but the process halts once dust grows to ~cm-sized pebbles due to the radial drift \cite{Weidenschilling1977} as well as the bouncing and fragmentation barriers \citep{Blum&Wurm2008}. \citet{Safronov1969} and \citet{Goldreich&Ward1973} independently proposed a shortcut to this approach, where dust settling into a gravitationally unstable thin layer leads to direct formation of planetesimals. However, it was soon found out that this settling causes vertical shear which triggers turbulence at a level that prevents the formation of the dense thin dust layer \citep{Weidenschilling1980}.

The current understanding of planet formation relies on dust instabilities as a means to increase dust densities and form self-gravitating clumps \citep{Chiang&Youdin2010}. The Streaming Instability (SI) \citep{Youdin&Goodman2005, Youdin&Johansen2007, Johansen&Youdin2007} is widely seen as the favourite mechanism to accomplish this. While simulations with single species show the SI is capable of producing high density dust clumps \citep{JohansenEtal2007}, multi-species SI with a dust size distribution seem to face some challenges. \cite{KrappEtal2019} found linear growth rates for discrete multi-species SI that are much lower than single species. \cite{PaardekooperEtal2020} showed that the SI vanishes in parts of the wave number space when approaching the continuous size distribution limit. In the nonlinear regime, \cite{Yang&Zhu2021} found that growth is still possible for multi-species SI when the dust to gas ratio $\mu$ exceeds unity \citep[see also][]{MatthijsseEtal2025}.

The Resonant Drag Instability (RDI) framework was recently put forward as a way to derive a family of instabilities that occur at low dust to gas ratio $\mu$ \citep{Squire&Hopkins2018a,Squire&Hopkins2018b}. RDIs develop when the gas supports a wave with a phase velocity that matches the (projected) streaming velocity of dust particles through the gas. This resonance leads to linear instability, often causing fast growth of the dust density perturbations. The choice of gas wave and dust streaming determines the type of RDI: sound waves lead to the Acoustic RDI \citep{Squire&Hopkins2018a,MagnanEtal2024a}, while inertial waves lead to either the streaming or settling instability \citep{Squire&Hopkins2018b,MagnanEtal2024b}, depending on the direction of the dust drift. The newly identified Dust Settling Instability (DSI) was found to grow much faster in the linear regime and required lower initial $\mu$ than the SI \cite{Squire&Hopkins2018b,Zhuravlev2019}.

\cite{Paardekooper&Aly2025a,Paardekooper&Aly2025b} generalised the RDI framework to polydisperse dust distributions with finite widths. They found that the resonance contribution to the instability is negative in the case of the SI but positive for the DSI. This means that in the linear regime, the low $\mu$ SI has no polydisperse counterpart in the limit of a continuous dust distribution \citep[but see][for an analysis of discrete multi-species dust that found a weak non-converging version of the SI]{KrappEtal2019}, whereas there is a polydisperse DSI with growth rates comparable to (but lower than) the monodisperse DSI. \cite{KrappEtal2020} found DSI linear growth rates that converge with the number of species, as opposed to the linear SI \citep{KrappEtal2019}. However, \cite{KrappEtal2020} found the nonlinear evolution of the mondisperse DSI to be insufficient: species with low dimensionless stopping time $ \taus \leq 0.01$ showed very little dust concentration (where $\taus$ is scaled by the orbital frequency $\Omega$), while species with $\taus = 0.1$ produced dust structures that were not dense enough for clumping and required timescales longer than the settling time. Their nonlinear numerical simulations treated every species individually, so the nonlinear polydisperse DSI remains largely unprobed, which is one of the main goals of this paper.

In this work we first of all revisit the nonlinear evolution of the monodisperse dust, the level of dust compression produced, and its convergence with spatial resolution. We then address the nonlinear evolution of a polydisperse dust distribution, its convergence with the number of species used to probe the distribution, and highlight how this treatment produces different density structures than numerical simulations with individual dust species. We start by describing our numerical setup in Section~\ref{section:setup}, followed by a presentation of the monodisperse (Section~\ref{section:mono}) and polydisperse (Section~\ref{section:poly}) results, before discussing our results in Section~\ref{section:discuss}.

\section{Setup}
\label{section:setup}
\subsection{Governing equations}
We perform a suite of 2D (R-Z) unstratified shearing box simulations \citep{Goldreich&Lynden-Bell1965,HawleyEtal1995} to study the nonlinear evolution of the DSI. Assuming Epstein drag \citep{Epstein1924}, the conservation equations in the local frame for a continuum of dust species \citep{PaardekooperEtal2021,Paardekooper&Aly2025a,Paardekooper&Aly2025b}:
\begin{align}
\partial_t\sigma 
+ \nabla\cdot(\sigma \ud )=& 0,\label{eq:dustcont}\\
\partial_t\ud
+ (\ud\cdot\nabla)\ud
=&
z_0\Omega^2 {\bf \hat z}- 2\bm{\Omega}\times {\bf u}+3\Omega^2 x {\bf \hat x}  - \frac{\ud - \vg}{\taus},\label{eq:dustmom}
\end{align}
where $\vg$ is the gas velocity, $\sigma$ is the size density \citep{PaardekooperEtal2020} and ${\bf u}$ is the size-dependent dust velocity, such that
\begin{align}
    \rho_{\rm d}=&\int \sigma \rmd \taus,\\ \rho_{\rm d}\bf{v_\rmd}=&\int \sigma \ud \rmd \taus,
\end{align}
with $\bf{v}_\rmd$ the bulk dust velocity and $\rho_{\rm d}$ the dust volume density.

The first term on the right hand side of (\ref{eq:dustmom}) is the dust vertical acceleration meant to account for stratification, with $z_0$ being the height below the midplane \citep[see][]{KrappEtal2020}. The second and third terms are the inertial forces due to the local frame, and the last term on the right hand side is the drag force acting on the dust. The governing equations for the gas read:
\begin{align}
\partial_t \rho_{\rm g} + \nabla\cdot (\rho_{\rm g} {\bf v}_{\rm g})=&0,\label{eq:gascont}\\
\partial_t\vg + (\vg\cdot\nabla)\vg =& -\mu z_0\Omega^2 {\bf \hat z} - 2\bm{\Omega}\times {\bf v}_{\rm g} +3\Omega^2 x {\bf \hat x} \notag \\&  + 2\eta {\bf \hat x} - \frac{\nabla p}{\rhog} + \frac{1}{\rhog}\int \sigma \frac{\ud - \vg}{\taus}\rmd\taus\,
\label{eq:gasmom}
\end{align}
where $\rhog$ and ${\bf v}_{\rm g}$ are the gas density and velocity, and $\mu$ is the total dust to gas density ratio. In order to establish hydrostatic equilibrium, the first term on the right hand side is introduced to balance the back-reaction of the vertical settling of dust particles \cite{KrappEtal2020}. We note that this offers a treatment of vertical stratification that is not self-consistent: stratification is only partly included through the $z_0\Omega^2 {\bf \hat z}$ term in the dust momentum equation. Gas stratification through the vertical density and pressure gradients is not included, but balance with dust vertical stratification is established through the $-\mu z_0\Omega^2 {\bf \hat z}$ term in the gas momentum equation. The second and third term are again the fictitious forces resulting from the choice of reference frame. The fourth term is the acceleration resulting from a radial global pressure gradient, where $2\rhog\eta=-\partial p/\partial r$\footnote{Note that our definition of the pressure support parameter $\eta$ is consistent with \cite{PaardekooperEtal2020,PaardekooperEtal2021,MatthijsseEtal2025}, but is different from the definition widely used in the literature \citep{Youdin&Goodman2005}, such that $\eta = r_0 \Omega^2 \eta_\mathrm{YG}$. This definition of $\eta$ is dimensional by choice, and results in a length scale $\eta/\Omega^2$ independent of $r_0$, which does not play a role in a local framework. The timescale is defined by $\Omega^{-1}$.} and we have assumed an isothermal equation of state with a fixed sound speed such that $\eta/(c\Omega)=10^{-3/2}$. Note that the drag back-reaction on the gas (last term) is now an integral over all dust species. 

Since we can only consider a finite number of dust species in our numerical treatment (namely; n=5, 10, and 20), the integrals are replaced by a sum over the dust species. We choose the species using Gauss-Legendre quadrature, whereby the stopping times (nodes) and corresponding densities (weights) are given as the roots of the $n^{th}$ Legendre polynomial over the width of the range of stopping times. This approach usually converges to the continuum faster than choosing stopping times that are equidistant in log-space \citep[see][for a comparison between the two sampling methods in the case of the polydisperse streaming instability]{MatthijsseEtal2025}.

\subsection{Equilibrium initial conditions}
Our simulations start with a spatially constant dust and gas densities. Our initial conditions employ an MRN dust size distribution \citep{MathisEtal1977}, where the dust size density is distributed according to $\sigma \propto a^{3+\beta}$, with $a$ being the dust size and the index $\beta=-3.5$. This distribution is motivated by observations of the interstellar medium and is thought to occur when dust evolution is dominated by either coagulation or fragmentation cascade \citep{BirnstielEtal2011}. Time-independent solutions can be found for the above governing equations where both gas and dust velocities are independent of the $y$ and $z$ coordinates and only the $y$ components of velocity depend on the $x$ coordinate \citep[see][for a full derivation]{PaardekooperEtal2020}. The resulting equations for the gas and dust velocities read:
\begin{align}
\vgx
=&
\frac{2\eta}{\kappa}\frac{\mathcal{J}_1}{\left(1+ \mathcal{J}_0\right)^2 + \mathcal{J}_1^2},
\\
\vgy
=&
-\frac{3}{2}\Omega x -\frac{\eta}{\Omega}\frac{1 +  \mathcal{J}_0}{\left(1+ \mathcal{J}_0\right)^2 + \mathcal{J}_1^2},\\
u_x =& \frac{2\eta}{\kappa}
 \frac{\mathcal{J}_1 - \kappa\taus(1 +  \mathcal{J}_0)}{(1+\kappa^2\taus^2)(\left(1+ \mathcal{J}_0\right)^2 + \mathcal{J}_1^2)} ,\\
 u_y   =& -\frac{3}{2}\Omega x
- \frac{\eta}{\Omega}\frac{1 +  \mathcal{J}_0 + \kappa\taus\mathcal{J}_1}{(1+\kappa^2\taus^2)(\left(1+ \mathcal{J}_0\right)^2 + \mathcal{J}_1^2)},
\end{align}
with integrals
\begin{align}
\mathcal{J}_m = \frac{1}{\rhog}\int \frac{\sigma (\kappa\taus)^{m}}{1+\kappa^2\taus^2} \rmd\taus.
\end{align}
where $\kappa$ is the epicyclic frequency, which is equal to $\Omega$ for the Keplerian discs studied here. We note that these solutions simplitfy to \citet{NakagawaEtal1986} equilibrium velocities in the case of a single dust species and a Keplerian disc. The equilibrium vertical velocities are:
\begin{align}
\vgz=&0,
\\
u_z=&z_0 \Omega^2 \taus
\end{align}
where we recall that the backreaction from the dust vertical settling is balanced hydrostatically.

\subsection{Numerics}
We fix the computational domain dimensions to $L_x = L_z = 0.1 z_0$ and we take $z_0$ to be equal to the gas scale height $H_g$. This domain height was chosen so that we can recover large structures that may form due to the DSI, without exceedingly violating the assumption of uniform vertical density (non-stratification). For a disc in hydrostatic equilibrium, one expects a gas density (and in case of an isothermal disc, pressure) change across our domain at the level of $\simeq 10\%$. This would increase to a $\simeq 50\%$ change if one uses a vertical extent $5$ times higher, at which point an unstratified setup becomes difficult to justify. A square domain with uniform radial and vertical resolution is chosen so that we do not impose any artificial restrictions on the dust structure. We use the pressure support parameter $\eta = 10^{-3}$ and fix $\Omega=1$, which gives an isothermal sound speed $c=10^{-3/2}$. The pressure support parameter $\eta$ controls the global pressure gradient in a disc, and hence the strength of the dust radial streaming, which in turn affects the level of clumping due to the SI. In the case of the DSI, the streaming of dust is mostly vertical, and is hence expected to be less affected by the value of $\eta$ (our choice of $\eta$ in this paper is consistent with \cite{Squire&Hopkins2018a} and \cite{KrappEtal2020}). 

The (initially uniform) gas density is $\rhog=1$ in code units and dust to gas ratio $\mu=0.1$. While this is much higher than the typically assumed $\mu=0.01$, this provides an idealised numerical setup to assess whether the DSI can be important for planet formation under favourable conditions. Growth rates in the linear regime are $\propto \mu^{1/3}$ for the monodisperse DSI and $\propto \mu^{1/2}$ for the polydisperse DSI \citep{Paardekooper&Aly2025b}, so we expect a slower evolution for $\mu=0.01$ by a factor of a few. On the other hand, local enhancement of $\mu$ due to dust drift can enable the favourable conditions examined here, provided there is a mechanism that stirs dust to high altitude. We seed the instability by perturbing the initial gas and dust velocities using white noise with a magnitude of $0.05c$. We run all simulations for a period of $100\Omega^{-1}$, which is 10 times larger that the dust settling time at $\taus=0.1$, so as to ensure we reach saturation.

The simulations were performed with the code \textsc{Idefix} \citep{LesurEtal2023} on the  \cite{DHPC2024}. \textsc{Idefix} is a publicly available finite-volume Godunov code written for modelling astrophysical fluid flows. We use a second order total variation diminishing Runge-Kutta time integrator. For the gas, we use the HLLC solver to compute the inter-cell fluxes \citep{ToroEtal1994}. For the dust, we use a solver similar to that implemented in \textsc{Athena++} \citep{Huang&Bai2022}, which sets the flux to zero for diverging Left/Right normal velocities. Importantly, we found it essential to implement a flag that reverts to flat reconstruction of face values whenever the dust density becomes negative or the velocity jumps by more than a factor of 5 from the neighbouring cell, to prevent the code from crashing. 

\section{Monodisperse simulations}
\label{section:mono}

We list the parameters for our monodisperse settling instability simulations suite in Table~\ref{table::mono_parameters}. Simulations run1 to run4 represent our fiducial monodisperse parameters, with $\taus=0.1$ and $\mu=0.1$, but at 4 different spatial resolutions. In simulations run5 to run9 we fix the resolution and $\taus$ and $\mu$ to correspond to the same values for our fiducial polydisperse simulation with 5 dust species, providing a direct comparison that will be presented in the next section, along with run10 which uses a value of $\taus$ equivalent to the average of the 5 dust species.

\begin{table}
    \centering
    \caption{Parameters for the monodisperse simulations suite.   
}
    \begin{tabular}{lccc}
        \hline
        Simulation & $\taus$ & $\mu$ & $N_x \times N_z$ \\
        \hline
        \hline
        run1 & 0.1 & 0.1 & 2048 $\times$ 2048 \\
        run2 & 0.1 & 0.1 & 1024 $\times$ 1024 \\
        run3 & 0.1 & 0.1 & 512 $\times$ 512 \\
        run4 & 0.1 & 0.1 & 256 $\times$ 256 \\
        \hline
        run5 & 0.00124 & 0.00338 & 512 $\times$ 512 \\
        run6 & 0.00289 & 0.01042 & 512 $\times$ 512 \\
        run7 & 0.01 & 0.023 & 512 $\times$ 512 \\
        run8 & 0.03455 & 0.03598 & 512 $\times$ 512 \\
        run9 & 0.08057 & 0.0272 & 512 $\times$ 512 \\
        \hline
        run10 & 0.037  & 0.1 & 512 $\times$ 512 \\
    \end{tabular}
    \label{table::mono_parameters}
\end{table}

\begin{figure}
    \centering
    \includegraphics[width=\columnwidth]{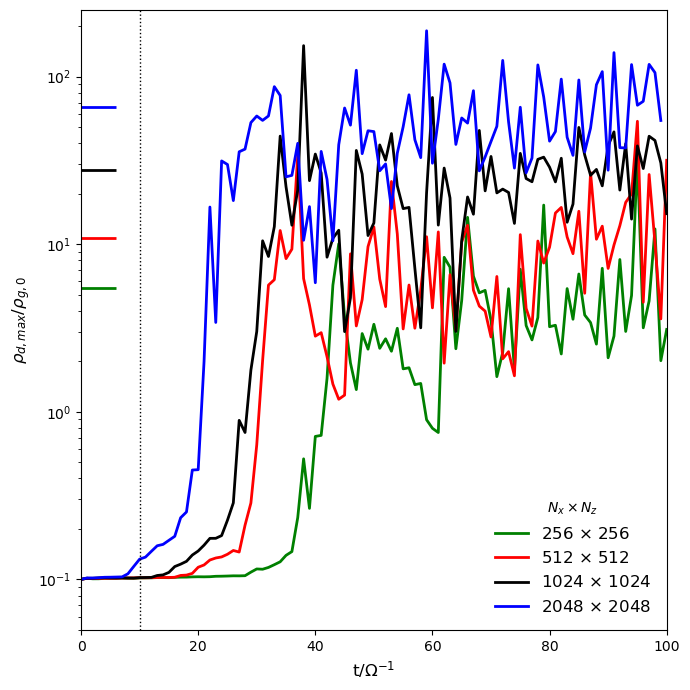}
    \caption{Evolution of the maximum dust density with time for the monodisperse settling instability with 4 different resolutions (run1 to run4 in Table~\ref{table::mono_parameters}). Short horizontal lines on the left indicate averages between $60$ and $100\Omega^{-1}$ and the dotted vertical line indicates the settling time.}
    \label{fig:mono_res_time_series}
\end{figure}

Figure~\ref{fig:mono_res_time_series} shows the time evolution of the maximum dust density for 4 different resolutions (run1 to run4). The maximum dust density clearly increases with resolution and the simulations do not yet show signs of convergence. We can also see that the instability is triggered and saturates earlier for higher resolutions; again with no signs of convergence. We note that this is a main difference between the settling and streaming instabilities: while the maximum dust density also increases with resolution for the streaming instability, the saturation time does not seem to get shorter, as it does for the settling instability \citep[see for example][Figure 7]{MatthijsseEtal2025}. This is particularly important since the main constraint on whether the settling instability contributes to planetesimal formation is the competition between the instability saturation and the settling timescales. At the highest resolution (run4), the monodisperse settling instability produces dust densities often exceeding $100\rhog$ (with a maximum of $188\rhog$ at $t=59\Omega^{-1}$, which slightly exceeds the Roche density \citep{Li&Youdin2021}, and an average of $66\rhog$ between $60$ and $100\Omega^{-1}$) and is thus likely to cause gravitational collapse. While the saturation time is still not faster than the settling time ($10\Omega^{-1}$ for $\taus=0.1$), it is likely to reach that value for a higher resolution run if the observed trend persists. 

\begin{figure*}
    \includegraphics[width=\textwidth]{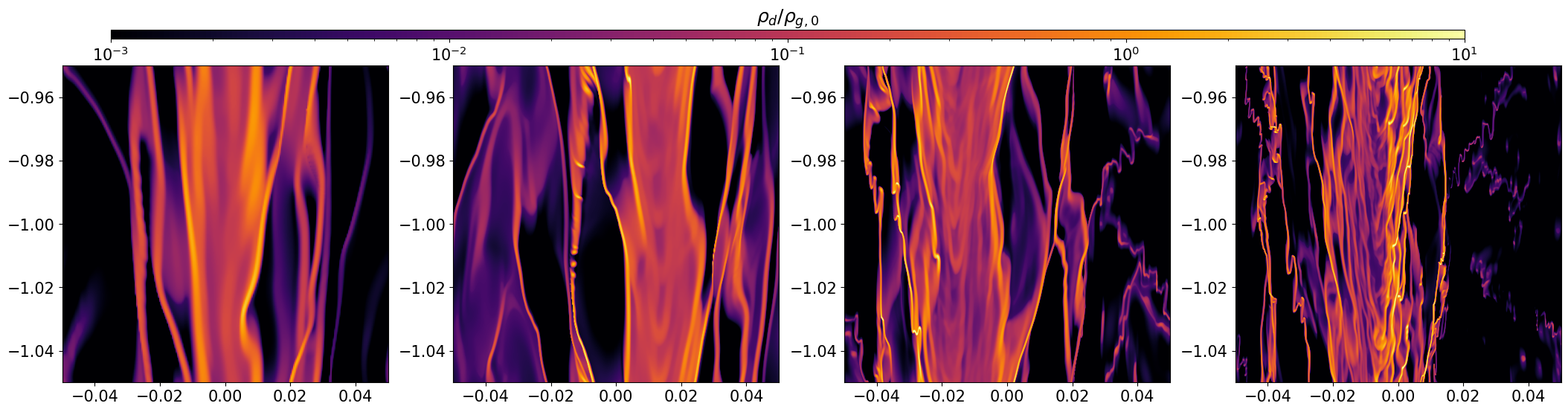}
    \caption{Dust density for the monodisperse simulations at $t=80\Omega^-1$. Panels show runs with different resolutions, increasing from left to right (run4 to run1).}
    \label{fig:mono_res_heatmap}
\end{figure*}

The instability can be visualised in Figure~\ref{fig:mono_res_heatmap} where we show dust density snapshots for simulations run1 to run4 at $t=80\Omega^{-1}$, with resolution increasing from left to right. The filamentary nature reported by \cite{KrappEtal2020} can be seen clearly in this figure. We also observe that as finer structure is resolved, higher dust densities are achieved within the filaments. This is likely to remain the case for resolutions higher than run4, which was the most computationally expensive simulation we performed. However, we note that we have not used any dust diffusion in these simulations. Dust diffusion will damp out modes at large wavenumbers, potentially leading to convergence with spatial resolution.

\begin{figure}
    \centering
    \includegraphics[width=\columnwidth]{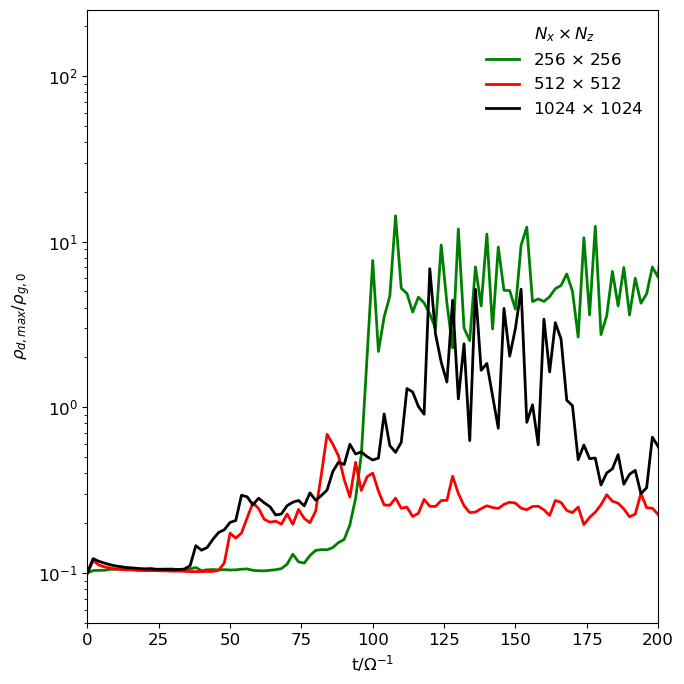}
    \caption{Same as Figure~\ref{fig:mono_res_time_series} but with FARGO3D and excluding the highest resolution (run1, blue curve). We use the same limits for the y-axis for an easier comparison, but we extend the x-axis to $t=200\Omega^{-1}$}
    \label{fig:fargo_time_series}
\end{figure}

\cite{KrappEtal2020} performed similar monodisperse simulations using the code FARGO3D \citep{Benitez-Llambay&Masset2016,BenitezEtal2019} and concluded that the settling instability is likely not important for clumping and planetesimal formation. To address this discrepancy, we ran simulations run2 to run3 with FARGO3D. The time evolution of the maximum dust density is plotted in Figure~\ref{fig:fargo_time_series} with the same colour code and y-axis limits as Figure~\ref{fig:mono_res_time_series} for an easier comparison (while extending the x-axis to $t=200\Omega^-1$). We notice that the green curve (run4) reaches comparable levels in both Figures (albeit with a slower triggering in the case of Figure~\ref{fig:fargo_time_series}, due to a smaller perturbation amplitude in the initialisation). However, going to higher resolutions, while the saturation time exhibits the same trend (earlier for higher resolutions), the maximum dust density obtained in the FARGO3D simulations becomes significantly lower, opposite to what we find in the \textsc{Idefix} simulations. We attribute this discrepancy to a numerical artefact that produces small scale stripes in the gas fields. These were reported in \citet[][section 3.5.5]{BenitezEtal2019} and \citet{MatthijsseEtal2025} in the case of the streaming instability, although there they do not seem to have an effect on dust. In the case of the settling instability, on the other hand, they do seem to negatively affect the dust density enhancement and this effect gets stronger at higher resolution. We explore this issue in some detail in Appendix \ref{app:fargo}. We note that \cite{KrappEtal2020} used a different shearing box dimension for their run corresponding to this $\mu$ and $\taus$ (same radial extent but five times larger in height), so a direct comparison is not intended here. However, in Appendix \ref{app:fargo} we repeat their simulation with the same box dimensions and obtain a maximum dust density consistent with their finding.

\begin{figure}
    \centering
    \includegraphics[width=\columnwidth]{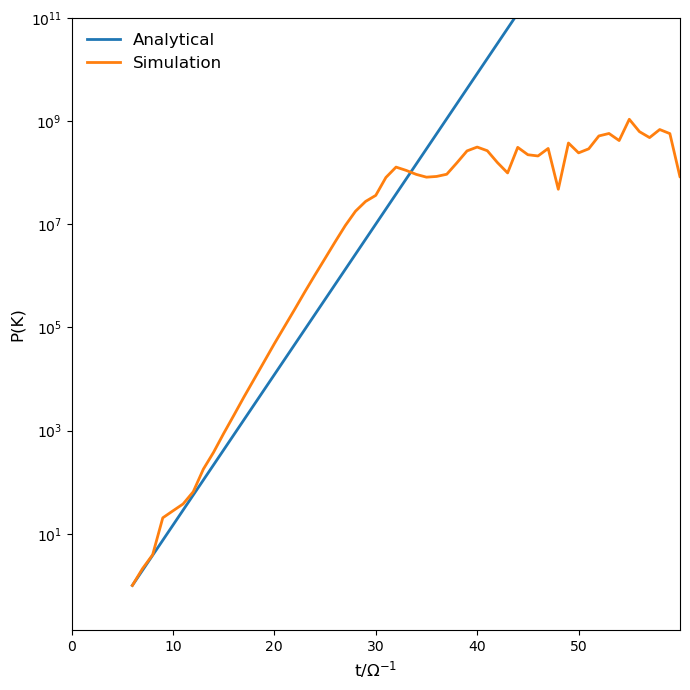}
    \caption{Amplitude of perturbation as a function of time for the single mode perturbation simulation (orange) compared to the amplitude of the same mode computed from the analytical growth rate.}
    \label{fig:singlemode}
\end{figure}

In order to make a connection with linear theory, we perform an additional simulation with the same setup as run2 (i.e, $1024^2$ grid cells), but with a single mode perturbation instead of white noise. We impose a small wave on the equilibrium gas and dust densities and velocities at unscaled wavenumber $K_x=31790$ and $K_z=1986$ with an amplitude of $10^{-4}$. We pick this particular wavenumber as it sits in the DSI instability region (i.e, close to the resonance) calculated from linear theory \citep[eg.]{Squire&Hopkins2018a, Paardekooper&Aly2025b} and the wave is resolved by $\simeq 32$ cells in the radial direction. In Figure~\ref{fig:singlemode} we compare the time evolution of the perturbation amplitude, computed from the absolute value of the Fast Fourier Transform of the dust density field, with the amplitude expected from the analytical growth rate $\Im(\omega(\mathbf{K}))$ calculated from linear theory. We find close agreement between the growth rates from our simulation and linear theory, until nonlinear evolution starts at $t\approx30\Omega^{-1}$. The  dust density evolution in the nonlinear state is very similar to that in run2 with the white noise perturbation, albeit with a somewhat later saturation (by about $10\Omega^{-1}$).

\section{Polydisperse simulations}
\label{section:poly}

\begin{table}
    \centering
    \caption{Parameters for the polydisperse simulations suite.   
}
    \begin{tabular}{lcccc}
        \hline
        Simulation & $\taus$ & $\mu$ & $N_x \times N_z$ & n\\
        \hline
        \hline
        poly5 & [0.001 - 0.1] & 0.1 & 512 $\times$ 512 & 5 \\
        poly10 & [0.001 - 0.1] & 0.1 & 512 $\times$ 512 & 10 \\
        poly20 & [0.001 - 0.1] & 0.1 & 512 $\times$ 512 & 20 \\
    \end{tabular}
    \label{table::poly_parameters}
\end{table}

After establishing the role of the DSI in forming high density dust structure in monodisperse simulations, we now turn to address how a distribution of dust sizes may change this picture. Our polydisperse simulations suite are summarised in Table~\ref{table::poly_parameters}. Here we fix the dust properties to $\mu=0.1$ and stopping times $\taus$ between 0.001 and 0.1 following MRN distribution \citep{MathisEtal1977}, with $\taus$ and $\mu$ for each species chosen as described in Section~\ref{section:setup}. We fix the spatial resolution to $512 \times 512$ grid cells, and change the number of dust species $\rm{n}=5,10$, and $20$. Our first objective is to determine whether the DSI converges with the number of dust species modelled, and if so, how different is the evolution from that of monodisperse dust. To this end we use our monodisperse run10 with $\taus=0.037$ for comparison, as this presents the average stopping time for our polydisperse simulations. We note, however, that since the monodisperse DSI is not very sensitive to the stopping time, a very similar comparison could be made using the monodisperse simulation run3 (with $\taus=0.1$).

\begin{figure}
    \centering
    \includegraphics[width=\columnwidth]{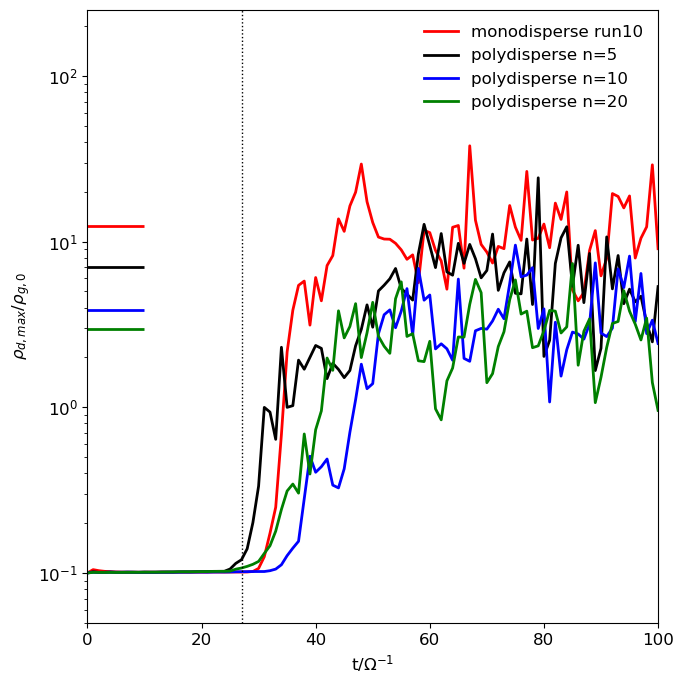}
    \caption{Evolution of maximum (total) dust density with time for 3 polydisperse simulations with $\rm{n}=5,10$, and $20$, as well as the monodisperse run10. The short horizontal lines to the left indicate averages between $60$ and $100\Omega^{-1}$ and the dotted vertical line indicates the settling time for the monodisperse run10 (with a stopping time $\taus=0.037$ equivalent to the average stopping time of the polydisperse runs).}
    \label{fig:poly_res_time_series}
\end{figure}

\begin{figure*}
    \centering
    \includegraphics[width=60.0mm]{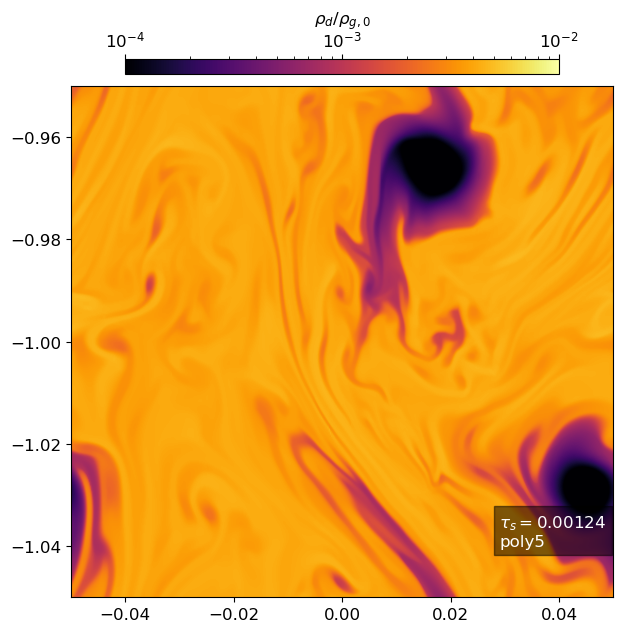}
    \includegraphics[width=60.0mm]{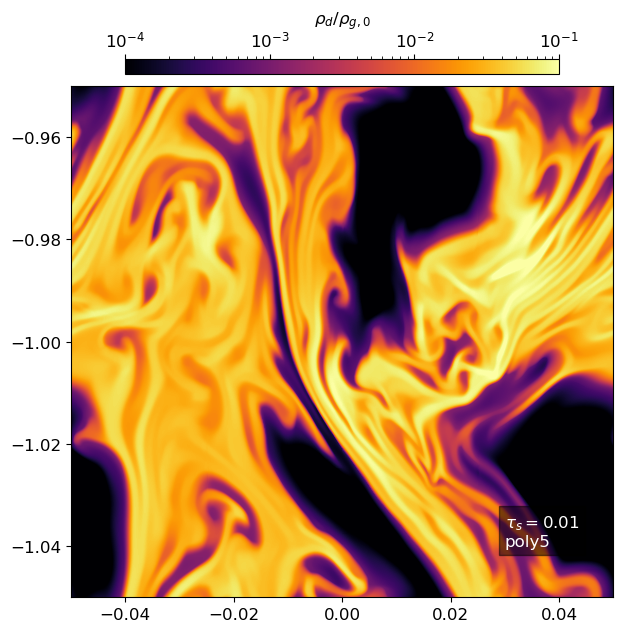}
    \includegraphics[width=60.0mm]{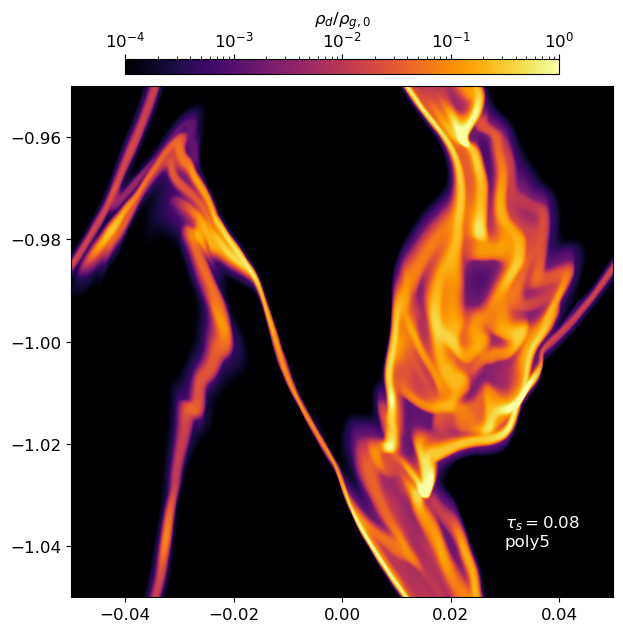}
    \includegraphics[width=60.0mm]{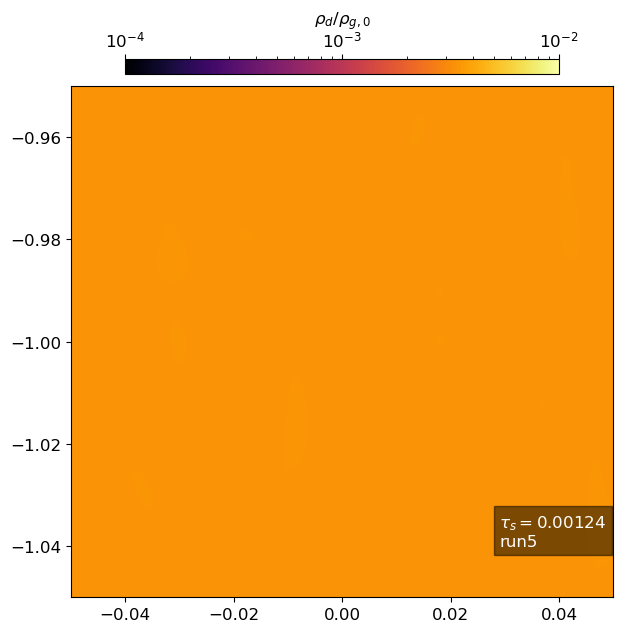}
    \includegraphics[width=60.0mm]{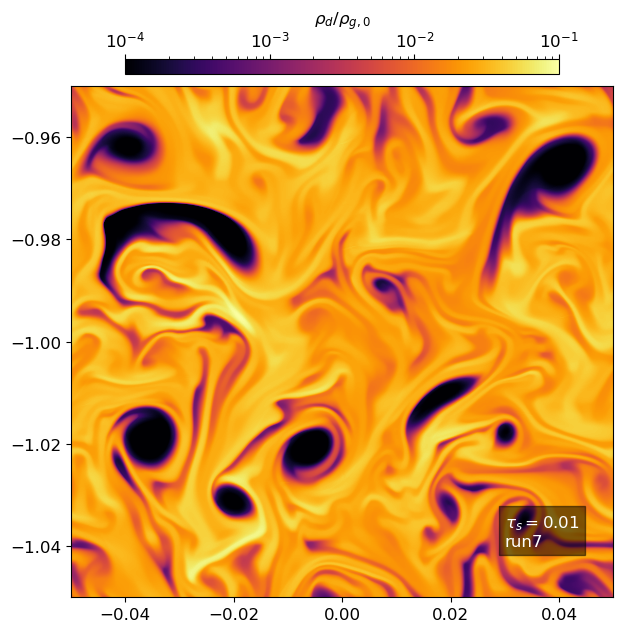}
    \includegraphics[width=60.0mm]{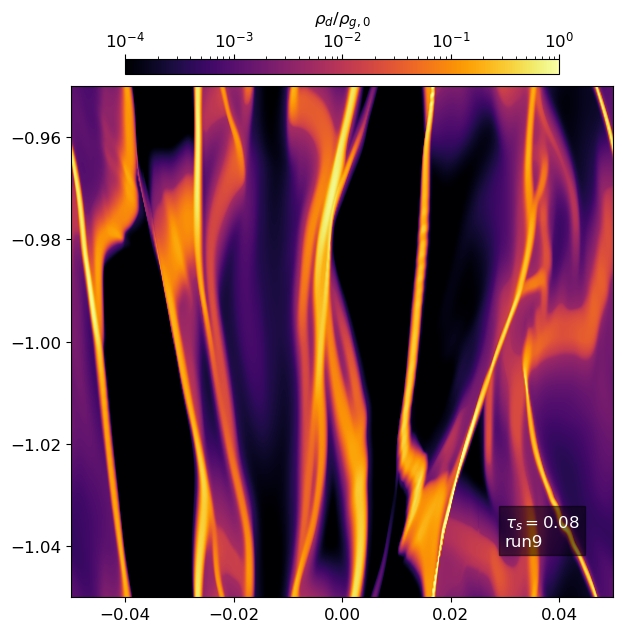}
    \caption{Dust density at $t=100\Omega^{-1}$ for 3 different species $\taus= 0.00124, 0.01$ and $0.08$ (left to right) for the polydisperse run poly5 (top panels) and the 3 corresponding monodisperse runs (bottom panels). Note that the colour bar limits are fixed for the same $\taus$ (top and bottom panels), but the maximum value increases by an order of magnitude as $\taus$ increases (left to right)}
    \label{fig:polyvsmono_heatmap}
\end{figure*}

In Figure~\ref{fig:poly_res_time_series} we plot the time evolution of the maximum (total) dust density for the three polydisperse simulations and with $\rm{n}=5,10$, and $20$, as well as the monodisperse run10. The short horizontal lines on the left of the plot indicate averages between $60$ and $100\Omega^{-1}$. The averaged maximum dust density for poly10 is much closer to poly20 than poly5, indicating signs of convergence with the number of dust species. We caution, however, that we fixed the spatial resolution for this comparison, which may mean this result is not general as spatial resolution may also affect the point at which we reach convergence with the number of species. One would need to vary both the spatial resolution and the number of dust species together for a more rigorous convergence study, which is beyond the scope of this work. We note that the maximum total dust density for poly20 is $~4$ times less than the monodisperse simulation with the same average $\taus$, a reduction that is similar to that in the polydisperse SI \citep{MatthijsseEtal2025}. In contrast to the SI, however, the saturation time for the DSI does not change much when going from mono- to polydisperse. This is crucial when assessing the viability of the polydisperse DSI, since the instability needs to develop within the settling time $1/(\taus \Omega)$. 

\subsection{Evolution of small dust: mono- vs polydisperse}
\cite{KrappEtal2020} looked at the nonlinear evolution of the DSI for different stopping times through multiple independent monodisperse simulations and concluded that the DSI produces very little density structure at low $\taus$ and $\mu$ (see their Figure 3). In Figure \ref{fig:polyvsmono_heatmap} we present the dust density maps for 3 different species $\taus= 0.00124, 0.01,$ and $0.08$ (left to right) for the polydisperse run (top panels) and the 3 corresponding monodisperse runs (bottom panels) for a snapshot at $t=100\Omega^{-1}$. These are the minimum, middle, and maximum stopping times for our poly5 simulation. While the DSI develops in the largest dust species to similar levels for both the mono- and polydisperse simulations, we can clearly see that for the intermediate and small dust species, the polydisperse treatment is able to capture much more density structure compared to the monodisperse simulations. We also observe a level of correlation between the structure formed in the different species in the polydisperse simulation.

\begin{figure}
    \centering
    \includegraphics[width=\columnwidth]{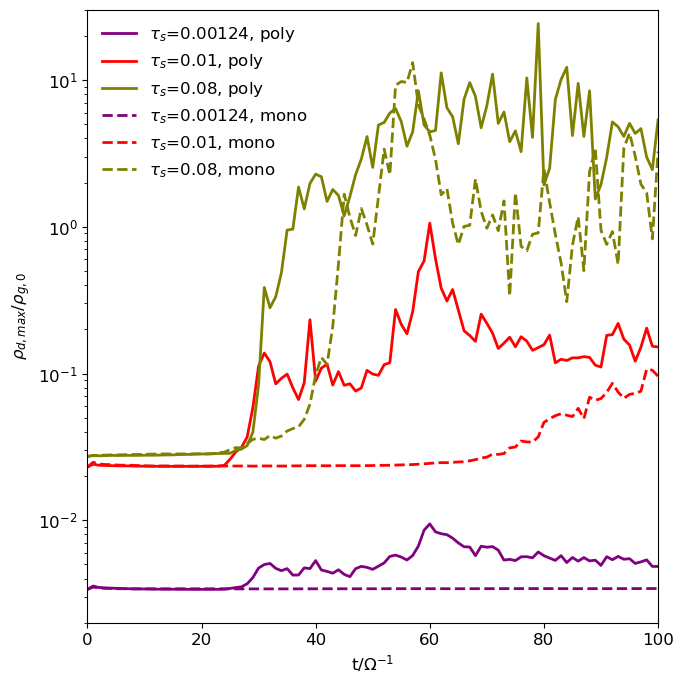}
    \caption{Time evolution of the maximum dust density for the same 3 species as in Figure~\ref{fig:polyvsmono_heatmap} for the monodisperse (runs 5, 7, and 9) and poly5 simulations.}
    \label{fig:polyvsmono_time_series}
\end{figure}

Figure~\ref{fig:polyvsmono_time_series} shows the time evolution of the maximum dust density for the same 3 species as in Figure~\ref{fig:polyvsmono_heatmap} for both the mono- and polydisperse simulations. Here we can appreciate more quantitatively the difference in evolution due to the polydisperse treatment. While the trigger time for the largest dust species is somewhat similar between poly- and mono-disperse simulations, we notice that the DSI is triggered earlier in the polydisperse simulation for the smaller dust species. We also see that, especially for the intermediate and small dust stopping times, the DSI results in larger dust density enhancements in the polydisperse simulations. Figure~\ref{fig:polyvsmono_time_series} indicate that, once the DSI is triggered for any dust species, its feedback on the gas causes the instability to occur in all other species. This is also hinted at by the apparent correlation observed in the structures formed by the DSI in the top panels in Figure~\ref{fig:polyvsmono_heatmap}.

\subsection{Morphology}

\begin{figure}
    \centering
    \includegraphics[width=\columnwidth]{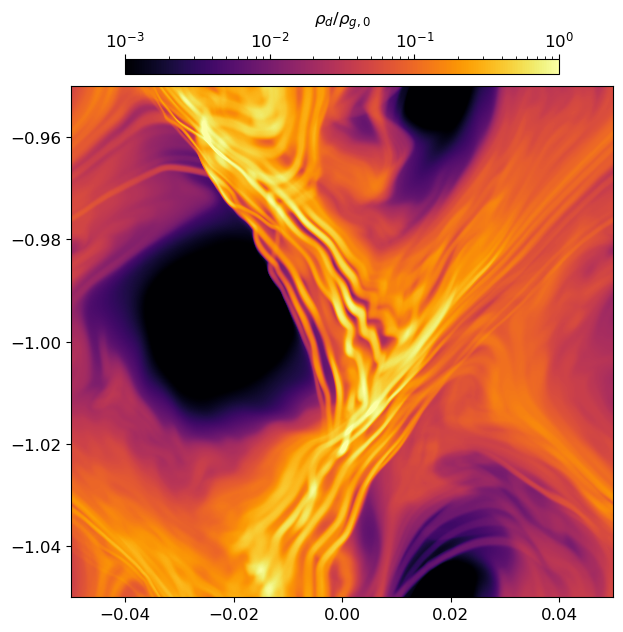}
    \caption{Total dust density for the poly20 run at $t=90\Omega^{-1}$.}
    \label{fig:poly20_heatmap}
\end{figure}

The correlation between dust structures at different $\taus$ shown in Figure~\ref{fig:polyvsmono_heatmap} motivates us to further investigate the morphology of the polydisperse simulations. In Figure~\ref{fig:poly20_heatmap} we show the \textit{total} dust density for the poly20 run at $t=90\Omega^{-1}$, i.e, the sum of the densities from all 20 dust species. The long, thin filaments each correspond to overdensities in different dust species towards the high end of the $\taus$ range. Hence, the large scale structure correspond to multiple smaller structures from the different species. Intriguingly, each species seems to form similarly-looking filaments to other species, but with a spatial offset. 

\begin{figure}
    \centering
    \includegraphics[width=\columnwidth]{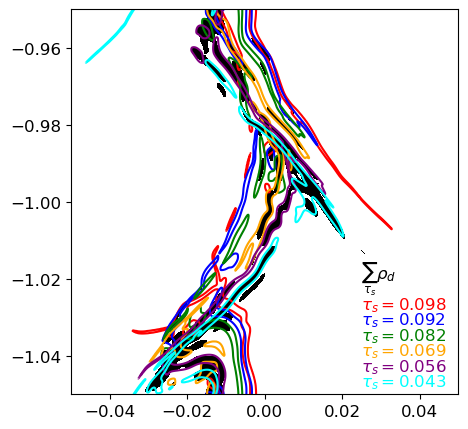}
    \caption{Coloured open curves represent contours of the top $2\%$ density for the highest $6$ stopping times (see figure legend) for the same snapshot shown in Figure~\ref{fig:poly20_heatmap}. Black filled curve represents the top $5\%$ total density of the sum of all 20 species.}
    \label{fig:poly20_contours}
\end{figure}

In Figure~\ref{fig:poly20_contours} we plot the top $2\%$ density for the highest $6$ stopping times (coloured open curves) as well as the top $5\%$ total density of the sum of all 20 species (black filled curve), for the same snapshot shown in Figure~\ref{fig:poly20_heatmap}. We notice how different dust species form different filaments and are coherent but offset from each other. The black filled contours, representing the highest total dust density regions, often coincide with separate filaments from individual species. While the structures formed in different dust species are clearly correlated, they tend to be spatially separated. This means that, while large scale structure will contain contributions from dust with different stopping times, at small scales dust interactions are mostly monodisperse; involving dust particles with similar stopping times. 

A closer inspection of Figure~\ref{fig:poly20_contours} shows the tendency for the size-segregated structures to form a $\taus$ gradient. The direction of the gradient varies with the details of the flow. The open contours in Figure~\ref{fig:poly20_contours} line up right to left in order of decreasing $\taus$ in the top multi-species filament, and the opposite direction in the middle filament. Examining the over-density in the sum of all species (black filled contours), we notice that they are more likely to overlap with lower $\taus$ contours (cyan and violet) than high $\taus$ (red and blue). The trend does not continue to even lower $\taus$; we only show the highest 6 out of 20 $\taus$ species to reduce clutter, but Figures~\ref{fig:polyvsmono_heatmap} and~\ref{fig:polyvsmono_time_series} show that low $\taus$ dust is less efficient in clumping.  

\begin{figure}
    \centering
    \includegraphics[width=\columnwidth]{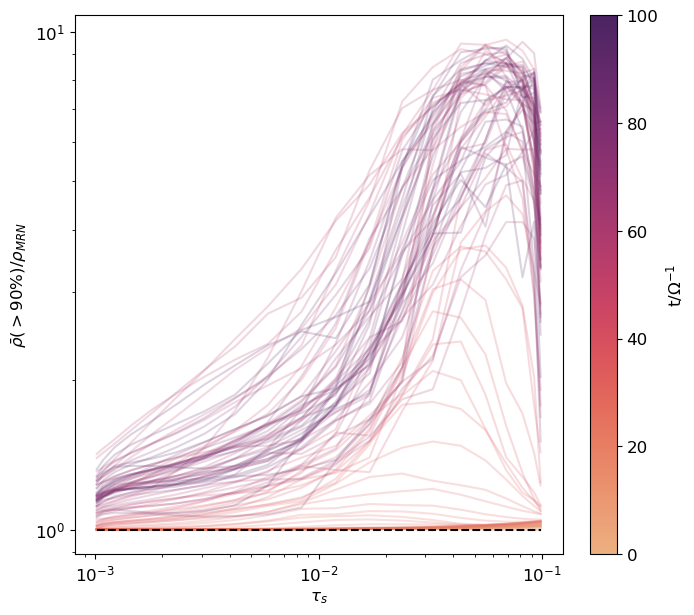}
    \caption{Evolution of the dust size distribution as a function of $\taus$ for the top $10\%$ in dust density, normalised by the initial MRN distribution.}
    \label{fig:size_distribution}
\end{figure}

The fact that the highest total density regions are more likely to overlap with structures formed by species that are not the high end of the $\taus$ range indicates that there might be a peak in the dust size distribution resulting from the DSI. This is similar to the peak reported in the polydisperse SI, which is also caused by the spatial segregation of dust with different size and the $\taus$ gradient that results from it \citep{MatthijsseEtal2025}. We test this implication in Figure~\ref{fig:size_distribution} where we plot the time evolution of the dust size distribution as a function of $\taus$ for the top $10\%$ in dust density, normalised by the initial MRN distribution. While there is a considerable scatter, we can see there is a tendency for the maximum of the distribution to occur at $\taus$ less than the maximum. Comparing Figure~\ref{fig:size_distribution} to Figure 6 in \cite{MatthijsseEtal2025}, we see that the SI produced a much better defined peak in the size distribution, with considerably less scatter, than the DSI. This may be due to the fact that the DSI tends to produce larger filaments stretching throughout the simulation box, and thus the statistics of the size distribution are heavily influenced by the evolution of each filament. Whereas the SI tends to produce numerous but smaller structures, and therefore the size distribution are less sensitive to evolution of each structure.

\section{Discussion and conclusion}
\label{section:discuss}

The DSI was shown to have analytical growth rates in the linear regime orders of magnitude higher than the SI and independent of the stopping time \citep{Squire&Hopkins2018a}. However, for the DSI to play an important role in planetesimal formation, it needs to reach maximum dust densities in the nonlinear regime comparable to those of the SI, and, importantly, form these high density structure faster than the settling time, which is seen as the more important barrier \citep{KrappEtal2020}. In our monodisperse simulations, our results show a trend suggesting that both criteria could be met at higher (currently unprobed) spatial resolutions, as higher resolution both increases the maximum dust density and shortens the time it takes for the DSI to form structure. This is a main difference between the SI and DSI: while spatial resolution increases the maximum dust density, it does not shorten the saturation time \citep{MatthijsseEtal2025}. This is likely due to the fact that linear growth in the DSI is dominated by the double resonance\footnote{At high radial wavenumber $k_x$ the growth rate in the linear regime increases without bound with wavenumber  \citep{Squire&Hopkins2018b}. This is called the double resonance angle as both the SI and DSI unstable branches merge (in wavenumber space), where the criterion ${\bf k}\cdot \Delta{\bf v} = 0$ is met (with $\bf v$ being the drift velocity).}, which increases indefinitely with wavenumber. We note, however, that the ability of the DSI to outpace the settling time is affected by the amplitude of the initial seed perturbations in our simulations. Regarding the maximum dust density attained by the DSI, diffusion (which was not taken into account in this paper) will set a limit on the maximum wavenumber that contributes to the DSI, and spatial convergence will be reached. Much like the SI \citep{Chen&Lin2020,McnallyEtal2021}, the level of dust diffusion and gas viscosity is also likely to limit clumping due to the DSI, as shown for the linear regime in Figure 10 in \cite{Paardekooper&Aly2025b}.

In this paper we attempt two types of comparisons between mono- and polydisperse DSI simulations. First, we compare polydisperse simulations with 5, 10, and 20 dust species (poly5, poly10, and poly20) to a single monodisperse simulation (run10) with dust to gas ratio $\mu$ equal to that of the total combined $\mu$ across all dust species in each polydisperse run, and a monodisperse stopping time equal to the average stopping time of the polydisperse runs (Figure~\ref{fig:poly_res_time_series}). This comparison showed that, while the maximum dust density obtained for the polydisperse runs is less than the monodisperse one by a factor of a few \citep[similar to the SI, see][]{MatthijsseEtal2025}, the saturation time is very similar between mono- and polydisperse runs (different to the SI trends). This is somewhat expected from the linear analysis in \cite{Paardekooper&Aly2025b}, where it is shown that the size distribution has a much more significant negative effect for the SI than the DSI. Figure~\ref{fig:poly_res_time_series} also showed that the polydisperse simulations seem to approach convergence with the number of dust species at about $n=10$ (with the caveat that this test was only performed at a fixed spatial resolution).

Second, we compare a single polydisperse simulation (poly5) with 5 different monodisperse runs (run5 to run9), each with $\mu$ and $\taus$ equal to a corresponding species in the poly5 run. This comparison showed that the polydisperse treatment results in a significantly different dust evolution. Structure forms for all $\taus$ in our size distribution in polydisperse simulations, whereas only high $\taus$ dust forms structure in corresponding monodisperse ones (Figures~\ref{fig:polyvsmono_heatmap} and~\ref{fig:polyvsmono_time_series}).

The spatial distribution of the highest density regions shows that the polydisperse DSI exhibits a filamentary structure. Dust with different sizes tend to form adjacent thin filaments, which seem to be ordered, with the largest dust forming filaments at the edge of the bigger structure (Figure~\ref{fig:poly20_contours}). The highest overall density regions tend to disfavour these edge filaments, therefore leading to a dust size distribution that is not monotonically increasing with dust size (Figure~\ref{fig:size_distribution}). \cite{MatthijsseEtal2025} recently reported a similar trend for the polydisperse SI, with the main difference that the SI forms several smaller filaments, whereas the DSI tends to form one bigger dust structure in spanning the computational domain.  

Our results indicate that the polydisperse DSI shows potential in forming clumps and is likely important for planetesimal formation. In cases where the DSI does not form clumps with high enough densities to trigger gravitational collapse (in a timescale shorter than the settling time), it can still play a role in creating overdensities that facilitate the onset of the SI or the closely related Vertically Shearing Streaming Instability \citep[VSSI,][]{Lin2021}. This latter scenario is particularly plausible as the DSI operates in the upper disc layers, whereas the SI and VSSI operate in the midplane as the dust settles, hence creating a natural ordering where the DSI leads to other drag instabilities, rather than compete with them.

\subsection{Caveats}

We note that an important caveat in this work is the local treatment of stratification, similar to \cite{KrappEtal2019}, represented by the terms proportional to $z_0$ in (\ref{eq:dustmom}) and (\ref{eq:gasmom}). Dust settling is inherently a stratified process and the DSI may manifest very differently if we take into account the vertical gas pressure gradient in a more realistic way.  
Moreover, this approximate treatment of vertical gravity obliged us to use a simulation box with small vertical extent (otherwise it becomes harder to justify the unstratified assumption), which may underestimate the resulting maximum dust density. We also had to add an effective acceleration to the gas to account for the vertical backreaction from the settling dust, which is strictly speaking only in equilibrium in the initial conditions and may inaccurately affect the evolution of both gas and dust. A more faithful assessment of the role of the DSI in planet formation requires a self-consistent stratified approach. Moreover, in such a self-consistent stratified vertically global setup, one would expect other instabilities to take place. These would include pure hydro instabilities like the Vertical Shear Instability \citep[VSI,][]{Urpin&Brandenburg1998,NelsonEtal2013} and the Kelvin-Helmholtz (KH) instability \citep{Sekiya1998,JohansenEtal2006}, as well as drag instabilities such as VSSI and the classical SI near the midplane or in regions where the dust to gas ratio approaches unity \citep{SchaeferEtal2025}. The interplay between these different instabilities may fundamentally change the nature of the DSI and its role in forming dust clumps in a stratified, vertically global setup.

We also note that even within the unstratified and vertically local treatment we adopted here, we have assumed a strictly isothermal equation of state and thus neglected the effects of cooling and buoyancy. A more careful treatment of the thermodynamics would implicitly take into account vertical gradients and shear, which is likely to affect both the VSI and DSI. \citet{Lehmann&Lin2023} performed linear stability analysis of a local unstratified shearing box but included the effects of buoyancy using the Boussinesq approximation \citep{Latter&Papaloizou2017} and showed that in the linear regime vertical buoyancy greatly stabilizes the DSI (apart from a narrow band in k-space around the resonances). However, they also reported a new instability, the Dust Settling Shearing Instability (DSSI), that can be thought of as a combination of the VSI and DSI when both dust-settling and vertical shear in the gas are accounted for, with growth rates in the linear regime that may exceed either of the classical versions of the instabilities. The interplay between these instabilities, and the way they manifest in the non-linear regime, warrants future investigation.


\noindent {\tiny \textit{Data Availability.} The data and scripts used to produce the figures in this paper are available at: \url{https://doi.org/10.4121/f45864ae-5e02-41e9-8a2f-f73a2d7086d7}}


\begin{acknowledgements}
We thank the anonymous referee for a detailed report and helpful suggestions. We thank Jip Matthijsse for fruitful discussions. This project has received funding from the European Research Council (ERC) under the European Union Horizon Europe programme (grant agreement No 101054502). The authors acknowledge the use of computational resources of the DelftBlue supercomputer, provided by Delft High Performance Computing Centre (https://www.tudelft.nl/dhpc).
\end{acknowledgements}

%
%

\bibliographystyle{aa}
\bibliography{refs}
\appendix
\section{Settling instability with FARGO3D}
\label{app:fargo}

\begin{figure}
    \centering
    \includegraphics[width=\columnwidth]{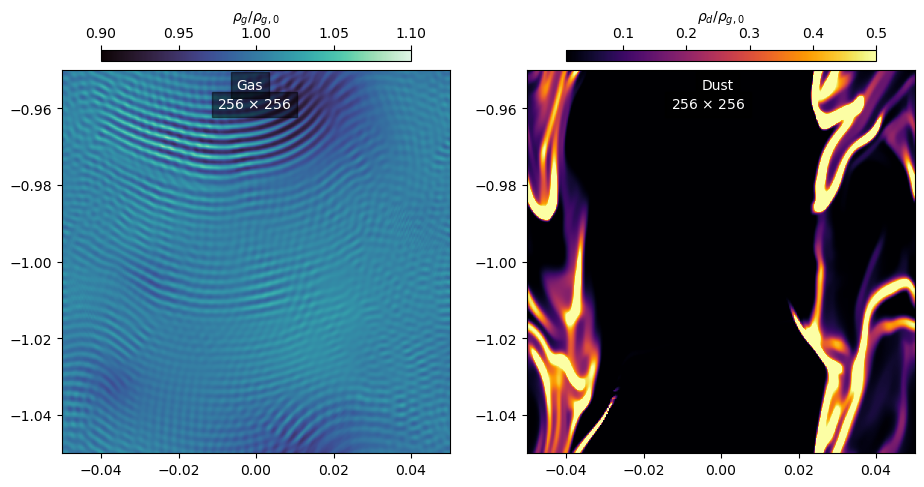}
    \includegraphics[width=\columnwidth]{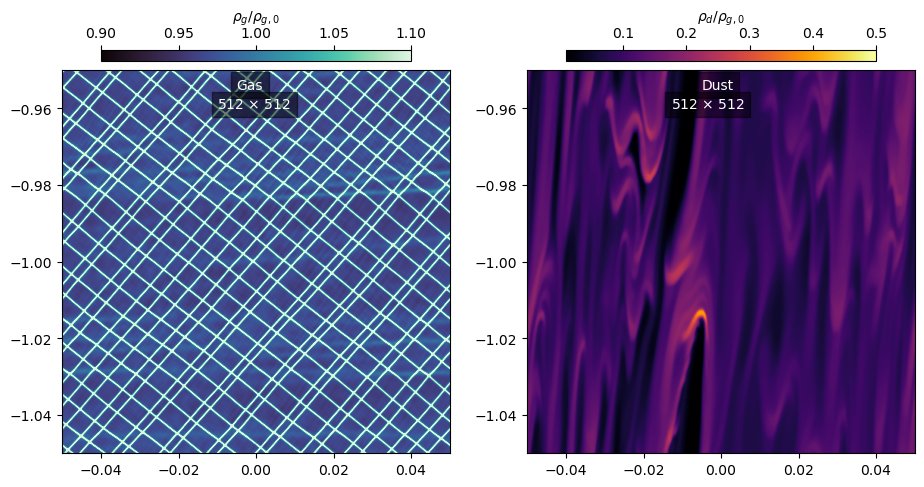}
    \includegraphics[width=\columnwidth]{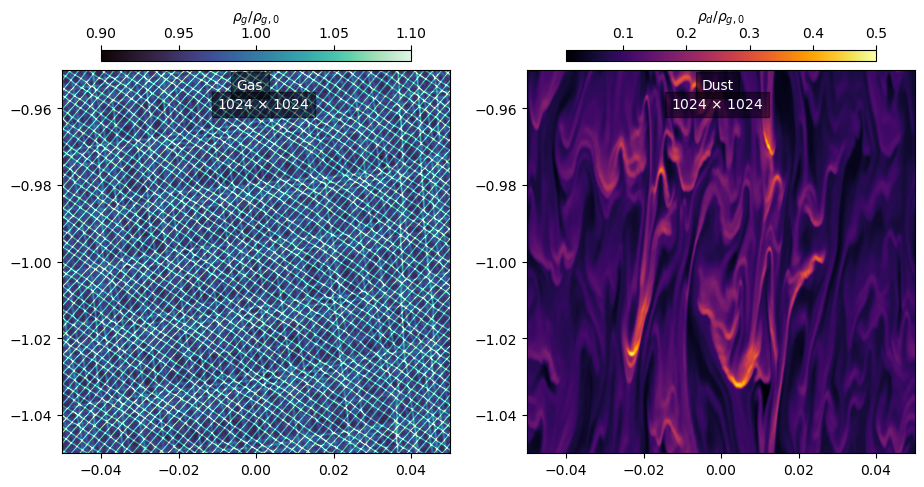}
    \caption{Gas (left) and dust (right) densities for a snapshot at $t=100\Omega^{-1}$ for run4, run3, and run2 (top to bottom). Note the visible stripes in the gas, while for the dust clumping is noticeably stronger for the lowest resolution}
    \label{fig:density_stripes}
\end{figure}

\begin{figure}
    \centering
    \includegraphics[width=\columnwidth]{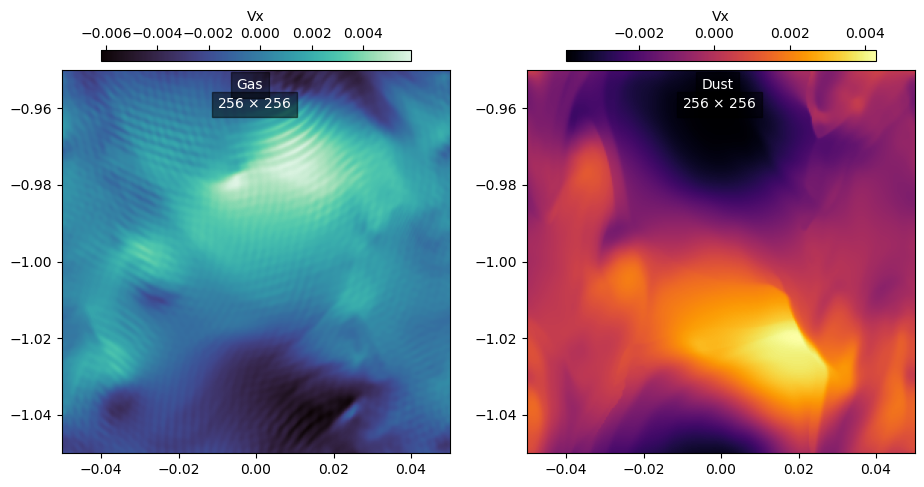}
    \includegraphics[width=\columnwidth]{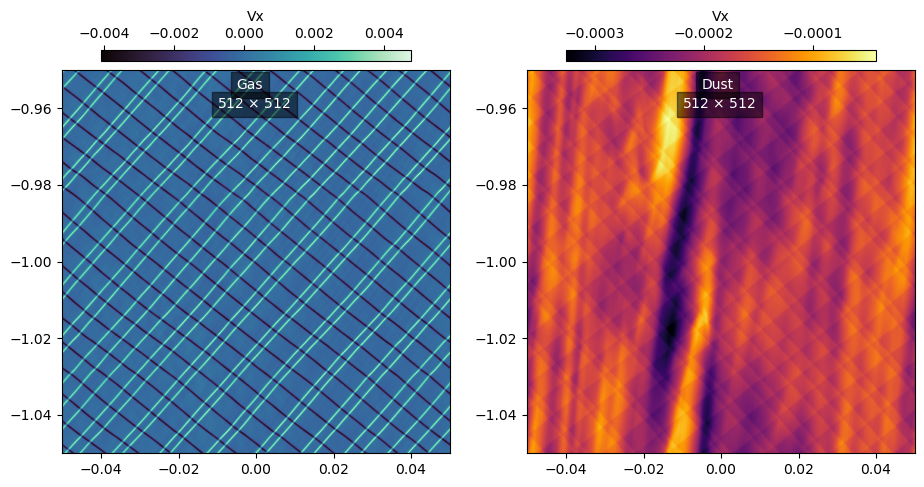}
    \includegraphics[width=\columnwidth]{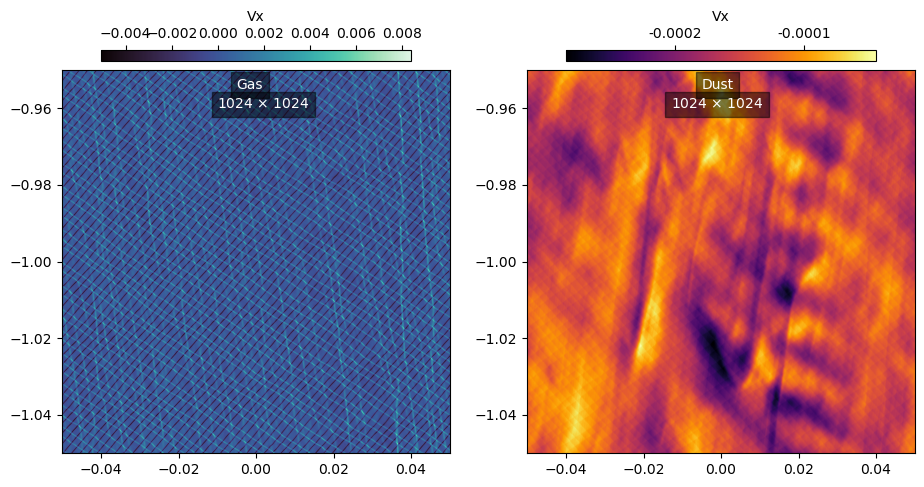}
    \caption{Same as Figure~\ref{fig:density_stripes} but for radial velocity. Here we can see the numerical stripes for run3 and run2 in both gas and dust}
    \label{fig:velocity_stripes}
\end{figure}

Our attempts to simulate the DSI with FARGO3D presented in Figure~\ref{fig:fargo_time_series} show a decrease in the dust ability to clump with increasing resolution from run4 to run3 (though it seems the trend is reversed increasing the resolution further from run3 to run2), which the opposite of what we expect. We attributed this to numerical artifacts that develop in the gas in the shape of stripes. These have been reported before in the case of the SI \citep{BenitezEtal2019, MatthijsseEtal2025}, in which case they have only been noticed in the gas and did not seem to affect the dust evolution. 

Here we take a closer look at this phenomenon. We plot the gas and dust density fields for a snapshot at $t=100\Omega^{-1}$ in Figure~\ref{fig:density_stripes} for run4, run3, and run2 (top to bottom). We can see clear stripes showing in the gas density for the higher resolution runs (run4 at the top only shows fainter ripple). While the stripes do not show in the dust density field, they seem to have a strong effect on the clumping, as it greatly decreases for the higher resolution runs where the gas stripes are more evident (note that while the color bar limits are different for gas and dust, they are kept the same for the different resolutions for an easier comparison). In Figure~\ref{fig:velocity_stripes} we show the radial velocity fields for both gas and dust for the same snapshot. Here we see the stripes in both gas and dust for the higher resolution runs, which strongly suggests that the negative effect on clumping at higher resolution is indeed caused by these numerical artifacts. Figures~\ref{fig:density_stripes} and~\ref{fig:velocity_stripes} also indicate a slight reduction of the strength of these stripes going from run3 to run2, which would be expected from the time series of the maximum dust density in Figure~\ref{fig:fargo_time_series}.

\begin{figure}
    \centering
    \includegraphics[width=0.48\columnwidth]{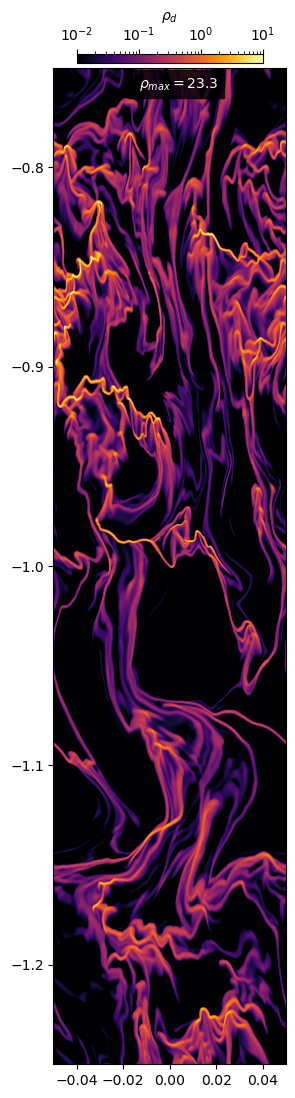}
    \includegraphics[width=0.48\columnwidth]{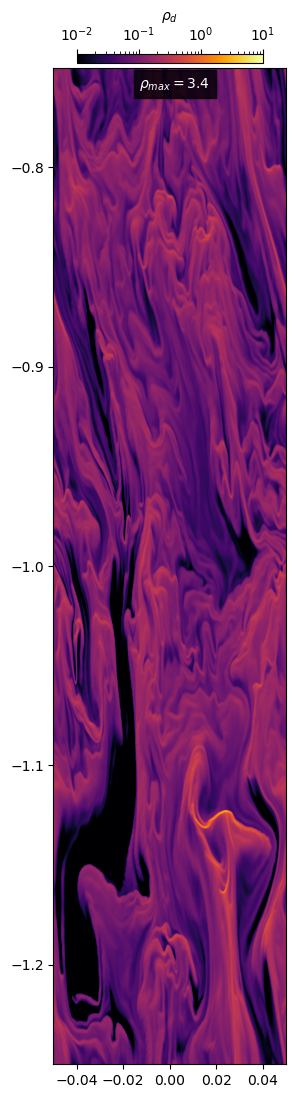}
    
    \caption{Dust desnity at $t=80\Omega^{-1}$ for a more vertically extended shearing box. Left: $1024 \times 2560$ grid cells, consistent with \cite{KrappEtal2020}. Right: $1024 \times 5120$ grid cells}
    \label{fig:krapp2020setup}
\end{figure}

\cite{KrappEtal2020} simulated the monodisperse DSI with varying physical numerical parameters using FARGO3D. Their simulation corresponding to our runs 1 through 4 (i.e, $\taus=0.1$ and $\mu=0.1$) used a box with dimensions $L_x = 0.1 z_0$ and $L_x = 0.5 z_0$ with $1024 \times 2560$ grid cells and reported a time-average of the maximum dust density $~22\rhog$. Note that this setup has higher resolution in the horizontal than in the vertical direction. We repeat this setup for a more consistent comparison. We also simulate the same box with higher vertical resolution with $1024 \times 5120$ grid cells to see whether the same trend observed in runs 2 to 4 applies here. We show a snapshot of the dust density at $t=80\Omega^{-1}$ for both simulations in Figure~\ref{fig:krapp2020setup}. The lower resolution run (left) is consistent with that reported in \cite{KrappEtal2020}. We notice that dust clumps more efficiently in the lower resolution run, with a maximum dust density of $23.3\rhog$, similar to that reported in \cite{KrappEtal2020}. The higher resolution run (right) shows a much less efficient clumping, with a maximum dust density almost an order of magnitude lower.

\end{document}